\documentclass[12pt]{article}
\usepackage{graphicx}
\usepackage{amsmath, amsfonts}
\usepackage{enumerate,array}
\usepackage[round]{natbib} 
\bibpunct{(}{)}{;}{a}{}{,}

\usepackage{url}
\usepackage[ruled]{algorithm2e}

\newcommand{\blind}{1}

\date{}

\begin{document}

\if1\blind
{
  \title{\bf  Fast approaches for Bayesian estimation of size of hard--to--reach populations using Network Scale-up} 
  \author{Leonardo S Bastos$^1$, 
Natalia S Paiva$^2$,  
Francisco I Bastos$^3$,
Daniel A M Villela$^1$ \\ \\
    $^1$Scientific Computing Program (PROCC), Oswaldo Cruz Foundation, Brazil \\
    $^2$National School of Public Health (ENSP), Oswaldo Cruz Foundation, Brazil \\
    $^3$Institute of Scientific and Technological Communication and Information in \\ Health (ICICT), Oswaldo Cruz Foundation, Brazil}
  \maketitle
} \fi

\if0\blind
{
  \bigskip
  \bigskip
  \bigskip
  \begin{center}
    {\LARGE\bf  Fast approaches for Bayesian estimation of size of hard--to--reach populations using Network Scale-up}
\end{center}
  \medskip
} \fi

\begin{abstract}
The Network scale-up method is commonly used to overcome difficulties in estimating the size of hard-to-reach populations. 
The method uses indirect information based on social network of each participant taken from the general population, but in some applications a fast computational approach would be highly recommended.
We propose a Gibbs sampling method and a Monte Carlo approach to sample from the random degree model.
We applied the abovementioned analytical strategies to previous data on heavy drug users from Curitiba, Brazil.  

\end{abstract}


\section{Introduction}

%
Hard-to-reach populations or groups are segments of a given society or community that for several different reasons such as stigma, marginalization and/or criminalization tend to hide and to avert interactions with formal institutions or people that they may perceive as being judgmental about their habits, attitudes or identities.
Overall, due to the very nature of their stigmatized behaviors and practices, those populations are not properly targeted/studied by classic sampling methods, either for the sake of contacting their members (for instance, to offer them help or social and medical assistance) or for the broad aim to estimate their size and to better understand who they are and where they tend to congregate and, especially, from whom and where they tend to evade or hide.

The spread of sexually transmitted infections in which stigmatized populations in some settings (e.g. gay, lesbian and transgender people in different societies) may play a core role for the spread of blood-borne infections driven by illicit (and in most places criminalized) habits such as injection drug use are frequently hard to fully understand, quantify and curb. 
Without fully understanding their extent and the spread of pathogens via social networks which composition and dynamic are largely unknown, evidence-based policies aiming to curb the epidemic may be very hard or sometimes impossible to formulate and implement. 
Therefore, knowing their sizes is quite important in order to design and evaluate public health policies to reduce the spread of a potential epidemic. \citep{bernard2010counting}.

Estimating the size of a population is an well-known problem in survey methods. 
However, they typically rely on information obtained from a question on whether a participant is a member of a population of interest.
For example, a direct question of the kind ``Are you a sex worker?''.
People in hard-to-reach populations, as part of stigmatized populations, tend to  hide their status, hence leading to underestimation of the size of the hard-to-reach population. 
\cite{Saletal11} illustrate this problem by estimating the number of heavy drug users in a Brazilian city using different methods, and they show that estimates from the direct approach are considerably lower than other estimates.

\cite{Beretal91} proposed a method, named Network scale--up (NSUM), to obtain information about hard-to-reach populations using the underlying contact network of the surveyed participants.
NSUM is based on indirect information about the hard-to-reach population, where surveys are conducted over the general population and people answer questions about contacts from this population. 
This might include: ``how many people do you know that are sex workers?''. 
Surveys also include questions inquiring about known contacts from well--known subsets of the general population, e.g., elderly people.
The underlying model is a Binomial model with an observed (surveyed) number of contacts from the hard--to--reach population, number of contacts from subgroups of known sizes, and unknown sizes for hard-to-reach populations.
%
%

In section 2, the Network scale-up method is presented where the random degree model is described.
%
%
In section 3, the inference procedure is introduced and a Gibbs sampling algorithm is proposed as an alternative to the Metropolis-Hastings algorithm proposed by \cite{maltiel2015estimating}. 
We also propose a Monte Carlo approach based on an approximation made in order to avoid using Markov chain Monte Carlo methods.
In section 4, we apply the random degree model with the proposed computational approaches in a network scale up study on heavy drug users in Curitiba, Brazil.
Finally, in section 5, we discuss limitations and future work.

\section{The Network scale-up method}

Let us suppose that we take a sample of $n$ individuals from the general population whose size is $N$. 
A questionnaire is applied and there are some questions of the type ``How many $X$s do you know that also know you, and you get in contact with each other in the last 12 months?''.
Here, $X$ represents any of several low frequency populations or subgroups, herein called populations for simplicity. A number $K$ of those populations have their size fully known, typically from official statistics. Conversely, a number $U$ populations are the hard-to-reach ones whose size is unknown and we intend to estimate.
%
%
 
Let $Y_{iu}$ be the number of people of a hard-to-reach population $u$ known by an individual $i$, $u=1,2,\ldots,U$, and $i=1,2,\ldots,n$. 
Let $X_{ik}$ be the number of people the individual $i$ know from population $k$, $k=1,2,\ldots,K$. 
Each individual is sampled from the general population according to a complex survey design and all the sampling information is then summarized in a sampling weight $w_i$, usually given by the inverse of sampling probability.

The individual contact network size, $\delta_i$, also called network degree, is the total number of contacts or friends of individual $i$. 
We assume that the number of people known by the participant $i$ from each population follows a Binomial distribution with parameters given by the network degree and the frequency of that population in the general population. 
Hence, 
\begin{eqnarray} \nonumber
Y_{iu} & \sim & Binom(\delta_i, \theta_u), \quad i=1,2,\ldots,n, \quad u=1,2,\ldots,U, \\ \label{NSUM}
X_{ik} & \sim & Binom(\delta_i, \pi_k), \quad k=1,2,\ldots,K,
\end{eqnarray}
where $\{\theta_u\} = (\theta_1,\ldots,\theta_U)$ are the frequencies of the hard-to-reach populations,
$\{\pi_k\} = (\pi_1,\dots,\pi_K)$ are the frequencies whose sizes are known in the general population, $\{N_k\} = (N_1,N_2,\ldots,N_K)$ respectively.
And $\{\delta_i\} = (\delta_1, \delta_2, \ldots, \delta_n)$ are the network individual degrees.
The equation (\ref{NSUM}) is the statistical basis of the Network scale-up method proposed by \cite{Beretal91}, and extended by \cite{zheng2006many}, \cite{McCormick2010}, and \cite{maltiel2015estimating}.

We set a Beta prior distribution for $\theta_u$ with parameters $a_u$ and $b_u$, i.e.
\begin{equation} \label{theta}
\theta_u \sim Beta(a_u, b_u), \quad a_u>0, \quad b_u>0, \qquad u=1,2,\ldots,u.
\end{equation}
We assume non-informative prior distributions for $\theta_u$, unless otherwise stated, setting $a_u = b_u = 1$.


Since the network degree is unknown, in a Bayesian fashion a prior probability distribution should be set. 
\cite{maltiel2015estimating} assume a log-normal prior for $\delta_i$ as in \cite{Raftery1988}.
We, however, assume a Gamma probability distribution for $\delta_i$, as follows 
\begin{equation} \label{degree}
\delta_i \sim Gamma(c_i, d_i), \qquad i=1,2,\ldots,n.
\end{equation}
The hyperparameters $c_i$ and $d_i$ are chosen in order to represent our knowledge about the degree network size. 
The random degree model, equations (\ref{NSUM}-\ref{degree}) is completed when the frequencies $\pi_k$, in equation (\ref{NSUM}) are set to the population frequency for that population, i.e. $\pi_k = N_k / N, \, \forall k$.
The choice for a Gamma distribution for $\delta_i$, equation (\ref{degree}), because we are able to sample from an approximation of its full conditional distribution. 
This will be discussed in the next section.

%
%

\section{Inference} \label{inference}

The posterior distribution of ($\{\theta_u\}, \{\delta_i\}$) is proportional to the product of the likelihood, given in equation (\ref{NSUM}), and the prior distributions, given by equations (\ref{theta}) and (\ref{degree}).
Samples from the posterior distribution can typically be obtained from a Markov chain Monte Carlo (MCMC) algorithm \citep{gamerman2006markov}.

\subsection{Gibbs sampling}

Since the Beta distribution is a conjugate prior of the Binomial distributions, the full conditional of $\theta_u$ is also a Beta distribution as the following
\begin{eqnarray} \label{thetaU.full.gs}
\theta_u \mid \{ \delta_i \}, \mathbf{y}, \mathbf{x} & \sim  & Beta\left( \sum_{i=1}^{n} y_{iu} + a_u, \sum_{i=1}^{n} (\delta_i - y_{iu}) + b_u \right), \quad u=1,2,\ldots U,
\end{eqnarray}
where $\mathbf{y} = y_{1,1},\ldots,y_{1,U},\ldots,y_{n,1}, \ldots,y_{n,U}$ and $\mathbf{x} = x_{1,1},\ldots,x_{1,K},\ldots,x_{n,1}, \ldots,x_{n,K}$.

For the network degree, \cite{maltiel2015estimating} use a Metropolis within Gibbs algorithm, where samples from the full conditional of the degree network are obtained from random walk Metropolis-Hastings algorithm \cite{Hastings1970}.
In fact, this model is already implemented in an R package called \textit{NSUM} \citep{Maltiel2015}. 
This approach, however, requires the user to tune the proposal variances in order to get reasonable acceptance rates. 
Aiming at a Gibbs sampler, we assume a Gamma prior for the network degree, as in equation (\ref{degree}), and use the Binomial to Poisson approximation in the likelihood. 
Then, the full conditional distribution for each network degree, $\delta_i$, is a truncated Gamma distribution, i.e. 
\begin{equation} \label{delta.full.cond}
\delta_i \mid \{\theta_u\}, \mathbf{y}, \mathbf{x} \sim TruncGamma_{(M_i, \infty)} \left(\sum_{u=1}^{U}y_{iu} + \sum_{k=1}^{K} x_{ik} + c_i, \sum_{u=1}^{U}\theta_{u} + \sum_{k=1}^{K} \pi_{k} + d_i \right),
\end{equation}
where $\pi_k = \frac{N_k}{N}$, for $k=1,2,\ldots,K$, and $M_i = \max({x_{i1},\ldots x_{iK},y_{i1},\ldots y_{iU}})$ for $i=1,2,\ldots, n$, which is a reasonable value, since the network degree must be at least greater than the number of contacts known by the participant from any population.
In order to sample from this distribution we use the method described by \cite{Nadarajah2006}.
The proposed Gibbs sampling algorithm to get $M$ simulation draws from random degree model is described in Algorithm \ref{alg.GS}.

\begin{algorithm}[htb]
 Initialize $\{\theta_u^{(0)}\}$ and $\{\delta_i^{(0)}\}$ \;
 \For{$m$ in 1 to $M$}{
 	\tcp{update $\{\theta_u\}$}
   \For{$u$ in 1 to $U$}{
   		sample $\theta_{u}^{(m)}$ from $p( \theta_u \mid \{\delta_i^{(m-1)}\}, \mathbf{y}, \mathbf{x} ) $ given in equation (\ref{thetaU.full.gs})\;
   }
   \tcp{update $\{\delta_i\}$}
   \For{$i$ in 1 to $n$}{
   		sample $\delta_{i}^{(m)}$ from $p( \delta_i \mid \{\theta_u^{(m)}\}, \mathbf{y}, \mathbf{x} ) $ given in equation (\ref{delta.full.cond})\;
   }
 }
\caption{Gibbs sampling algorithm for the random degree model.} \label{alg.GS}
\end{algorithm}

\subsection{Monte Carlo approach}

In order to simplify the analysis and avoid using a MCMC method, we could assume that we learn about the network degree just using information from the populations whose sizes are known.
Therefore, the posterior distribution of $(\{\delta_i\}, \{\theta_u\})$ can be approximated by
%
%
\begin{eqnarray} \nonumber
p\left(\{\theta_u\}, \{\delta_i\} \mid \mathbf{y}, \mathbf{x} \right) & = & p\left( \{\theta_u\} \mid \{\delta_i\}, \mathbf{y}, \mathbf{x} \right) p\left( \{\delta_i\} \mid \mathbf{y}, \mathbf{x} \right)  \\ \label{mc.post}
& \approx & p \left( \{\theta_u\} | \{\delta_i\},  \mathbf{y}, \mathbf{x} \right) p \left( \{\delta_i\} \mid \mathbf{x} \right).
\end{eqnarray} 
The joint distribution of $[\{\theta_u\} | \{\delta_i\},  \mathbf{y}, \mathbf{x}]$ consists in independent Beta distributions given by
\begin{equation} \label{mc.thetaU}
\theta_u \mid  \{\delta_i\},   \mathbf{y},  \mathbf{x} \sim  Beta\left( \sum_{i=1}^{n} y_{iu} + a_u, \sum_{i=1}^{n} (\delta_i - y_{iu}) + b_u \right), \quad u=1,2,\ldots U.
\end{equation}
And the joint distribution of $[\{\delta_i\} \mid \mathbf{y}, \mathbf{x}]$ is given by independent truncated Gamma distributions as follows
\begin{equation} \label{mc.deltai}
\delta_i \mid  \mathbf{x} \sim TruncGamma_{(m_i, \infty)} \left(\sum_{k=1}^{K} x_{ik} + c_i, \sum_{k=1}^{K} \frac{N_{k}}{N} + d_i \right), \quad i=1,2,\ldots, n,
\end{equation}
where $m_i = \max({x_{i1},\ldots x_{iK}})$. 
Sampling from the joint posterior distribution, equation (\ref{mc.post}), is trivial. 
We need to sample first $n$ network degrees from (\ref{mc.deltai}). 
Then, conditional on the network degree generated we sample $U$ frequencies $\{\theta_u\}$ from distribution given in equation (\ref{mc.thetaU}).
The algorithm to obtain $M$ Monte Carlo samples from the distribution (\ref{mc.post}) is presented in Algorithm \ref{alg.MC}.
Notice that algorithm \ref{alg.MC} is not a MCMC algorithm.
\begin{algorithm}[htb]
   \tcp{Sampling $\{\delta_i\}$}
  \For{$i$ in 1 to $n$}{
	   sample $M$ values of $\delta_{i}$ from $p( \delta_i \mid \mathbf{x} ) $ given in equation (\ref{mc.deltai})\;
   }
   \tcp{Sampling $\{\theta_u\}$}
   \For{$m$ in 1 to $M$}{
		\For{$u$ in 1 to $U$}{
   			sample $\theta_{u}^{(m)}$ from $p( \theta_u \mid \{\delta_i^{(m)}\}, \mathbf{y}, \mathbf{x} ) $ given in equation (\ref{mc.thetaU})\;
		}
   	}
\caption{Monte Carlo approach for the random degree model. } \label{alg.MC}
\end{algorithm}

%
%
\section{Results}

\cite{Saletal11} estimated the total number of heavy drug users in Curitiba, Brazil in 2010 using the network scale-up method. 
A heavy drug user is defined by a person who use illicit drug other than marijuana approximately 25 times in the last 6 months (an average of once a week) after the definition proposed by PAHO/WHO (Pan American Health Organization/World Health Organization).
The dataset is household-based random sample of 500 adult (i.e., aged $\geq$18 years) residents of Curitiba, Brazil in 2010.
Besides the demographic questions, participants were asked how many people they know from several populations that also know the participant, and have been in contact in the last 2 years.
The dataset is freely available (upon registration) in a website of the Office of Population Research, Princeton University (\url{http://opr.princeton.edu/archive/nsum/}).

There are questions about the number of people in the participant's contact network that belongs to twenty populations whose size are known, such as public university students, women over 70 year-old, etc. 
And also the number of contact network friends that are members of four hard-to-reach populations:
heavy drug users, female sex workers (FSW), men who have sex with another men (MSM), and women who had an abortion in the last 12 months (Abortion).

For each approach, 5,000 initial iterations were run in order to know how many MCMC runs are needed according to Raftery and Lewis diagnostic implemented in coda R package \citep{Plummer2006}.
Also, the random effects model using the Metropolis-Hastings algorithm required no less than 60,000 runs, so we run 80,000 runs for all models. 
The computational time for 80,000 MCMC iterations and 80,000 samples from the Monte Carlo approach for the random degree are presented in Table \ref{tbl.comp.times}.
Even though a considerably smaller number of samples were already enough for the Monte Carlo approach we run 80,000 Monte Carlo samples for the sake of comparison.
The same idea applies to the Gibbs sampling approaches, since less than 10,000 simulations were enough to achieve convergence.
Convergence was checked using Raftery and Lewis, Gelman and Rubin and Geweke diagnostics \citep{Plummer2006}.
Analyses were performed on a Linux-mint desktop (Intel i7-4820K, 3.70 GHz, 8 cores).
\begin{table}[ht]
\caption{Running times (in minutes) for obtain 80,000 samples using different computational approaches. } \label{tbl.comp.times}
\begin{center}
\begin{tabular}{rlc}
  \hline
Abbreviation  & Model & Time (minutes) \\ 
  \hline
Metropolis   & Random degree model using Metropolis-Hasting       & 28.55 \\ 
Gibbs        & Random degree model using Gibbs sampling           & 23.63 \\ 
Monte Carlo  & Random degree model using the Monte Carlo approach &  1.29  \\ 
   \hline
\end{tabular} 
\end{center}
\end{table}

\begin{figure}[ht]
\includegraphics[scale=.75]{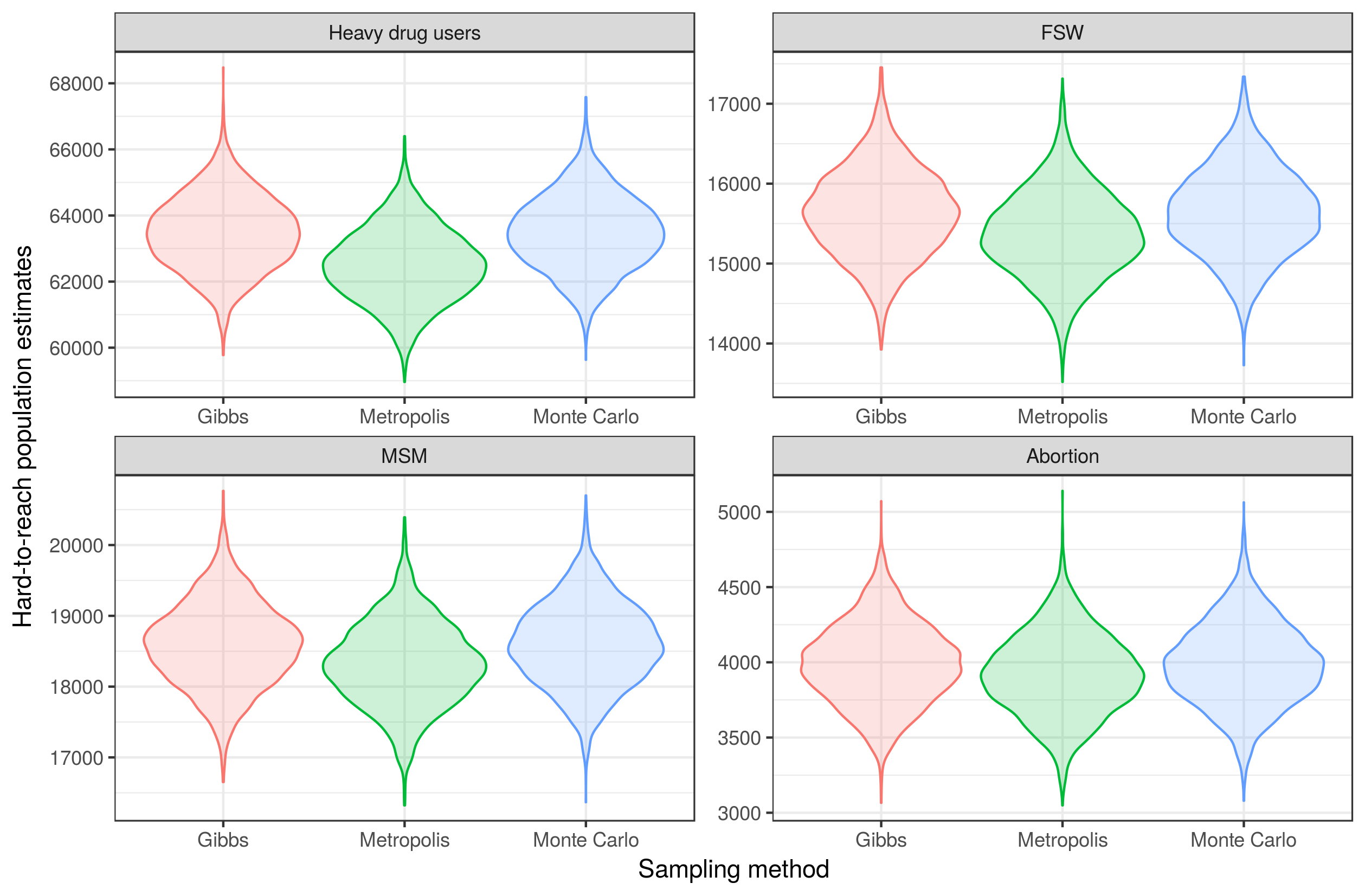}
\caption{Violin plots of the posterior distribution of the size of hard-to-reach populations in Curitiba, Brazil using different computational approaches.} \label{fig.h2r.estimates}
\end{figure}

An initial burn-in period containing 10,000 iterations was ignored for each one of the four chains, and to avoid autocorrelation we set a thinning parameter of 70.
Hence we end up with 1,000 samples for each chain. 
For the analysis, we generate 4,000 samples using the Monte Carlo approach, which demanded only 3.8 seconds in the same machine. 
The posterior estimates for the hard-to-reach populations in Curitiba are presented the violin plots are in Figure \ref{fig.h2r.estimates}.
The hard-to-reach population estimates differ from each other, but they are very similar among the different computational methods. 
The estimate for the total of heavy drug users in Curitiba is 63.5 thousand inhabitants (95\% CI 61.4k, 65.6k), corresponding to 3.5\% of the Curitiba's population. 
This result is consistent with \cite{Saletal11} and \cite{maltiel2015estimating} findings using the NSUM estimates and random degree model respectively.
We also found that in Curitiba there were approximately 15.6 thousands of female sex workers (95\% CI 14.6k, 16.7k), 18.5 thousands of men who had sex with other men (95\% CI 17.4k, 19.8k), and four thousand of women who did an abortion in the last twelve months (95\%CI 3.5k, 4.5k).


We also estimated that on average each person has approximately 200 network contacts, which is consistent with \cite{kudo2001neocortex} that claims there is a limited number of contacts/friends one could make social stable relationship. 
They claim this number of contacts should vary from 150 to 500.
The range of the posterior mean of the individual network degrees in Curitiba data goes from 5 to 2,000 network contacts.

%
%

\section{Discussion}

The NSUM can be applied in large national surveys where questions are asked about the number of the participant contacts who are members of known and unknown populations. 
We present a Gibbs sampling algorithm and a Monte Carlo approach to perform fast inference in the random degree model described in \cite{maltiel2015estimating}, used to estimate the size of hard-to-reach populations. 
The Monte Carlo approach is considerably faster than MCMC applying either Gibbs sampling or Metropolis, and the Monte Carlo estimates are quite similar to the ones from either of MCMC approaches. 
We illustrate the proposed method in a Network scale-up study on heavy drug users in Curitiba, Brazil. 

The Monte Carlo approach relies on an approximation that essentially does not distinguish the conditional distributions of the network degrees given the structure of the known population contacts from the unknown population contacts.  The Gibbs sampling approach shown here already has an advantage over a Metropolis-Hastings algorithm, since we can sample from the full conditional distribution of the network degree.  Whereas a large number is sampled from the distribution of degrees using Gibbs sampling, the Monte Carlo approach does sampling of the network degrees disregarding the structure of the network of hard--to--reach population.
Such approximation permits a dramatically faster estimation.

The NSUM suffers from different sources of bias, for example barrier effects and transmission bias \citep{Killworth2003, Killworth2006, Saletal11, maltiel2015estimating}. 
Barrier effects are present when some people may know more individuals with characteristics similar to themselves.
Transmission bias means that a person may not be aware that one of his network contacts belongs to a specific population, specially hard-to-reach populations. 
In this work, we assume no presence of such biases. 
However, if the assumption of absence of these biases is too strong, we recommend using the barrier effects and/or the transmission bias models proposed by \cite{maltiel2015estimating}, which depend on a Metropolis-Hastings algorithm.

Our estimations were very much similar when using Metropolis-Hastings, Gibbs sampling and the Monte Carlo approach for the network scale-up study illustrated here. Once it is known that barrier effects and transmission bias do not impact significantly on estimations, we could argue for using a fast Monte Carlo approach in studies of this kind.

\section*{Acknowledgement}

The authors have full or partial support of the following Brazilian funding agencies CAPES, CNPq and FAPERJ.

\bibliographystyle{harvard}
\bibliography{ref}

\end{document}